# Layer-dependent pressure effect on electronic structures of 2D black phosphorus

Shenyang Huang[1,+], Yang Lu[2,*,+], Fanjie Wang[1], Yuchen Lei[1], Chaoyu Song[1], Jiasheng Zhang[1], Qiaoxia Xing[1], Chong Wang[1], Yuangang Xie[1], Lei Mu[1], Guowei Zhang[3], Hao Yan[4], Bin Chen[2] & Hugen Yan[1,5*]

[1]State Key Laboratory of Surface Physics, Key Laboratory of Micro and Nano - Photonic Structures (Ministry of Education), and Department of Physics, Fudan University, Shanghai 200433, China

[2]Center for High Pressure Science & Technology Advanced Research, Shanghai 201203, China

[3]Institute of Flexible Electronics, Northwestern Polytechnical University, Xi'an 710072, China

[4]CAS Key Laboratory of Experimental Study under Deep-sea Extreme Conditions, Institute of Deep-Sea Science and Engineering, Chinese Academy of Sciences, Sanya 572000, China

[5]Collaborative Innovation Center of Advanced Microstructures, Nanjing 210093, China.

[+]These authors contributed equally to this work.

*E-mail: hgyan@fudan.edu.cn (H. Y.), yang.lu@hpstar.ac.cn (Y. L.)

**Through Infrared spectroscopy, we systematically study the pressure effect on electronic structures of few-layer black phosphorus (BP) with layer number ranging from 2 to 13. We reveal that the pressure-induced shift of optical transitions exhibits strong layer-dependence. In sharp contrast to the bulk counterpart which undergoes a semiconductor to semimetal transition under ~1.8 GPa, the bandgap of 2 L increases with increasing pressure until beyond 2 GPa. Meanwhile, for a sample with a given layer number, the pressure-induced shift also differs for transitions with different indices. Through the tight-binding model in conjunction with a Morse potential for the interlayer coupling, this layer- and transition-index-dependent pressure effect can be fully accounted. Our study paves a way for versatile van der Waals engineering of two-dimensional BP.**

Recently, the renaissance of black phosphorus (BP) as an emerging layered two-dimensional (2D) semiconductor has attracted tremendous research interests [1-3]. Different from graphene and transition metal dichalcogenides (TMDCs), BP has a direct bandgap ranging from 0.35 eV (bulk) to 1.7 eV (monolayer) [4,5], which, in conjunction with the remarkable in-plane anisotropy, makes BP a promising candidate for versatile opto-electronic devices [6-10].

Pressure can readily induce changes of lattice constants, especially the interlayer distance in layered 2D materials, hence modify the electronic properties of materials [11], such as semiconductor to metal transitions [12,13], bandgap opening in gapless few-layer graphene [14], and band structure engineering in van der Waals heterostructures [15] and Moire superlattices [16,17]. Previous studies show that pressure induces a semiconductor-to-metal transition at ~1.8 GPa on bulk BP [18-20], which suggests that the band gap shrinks with increasing pressure. It is tempting to infer that even for thin layers, the band gap also shrinks under pressure and eventually closes up to become a metal [21]. However, except for a few studies on pressure-induced lattice structure transition in thin layers [22-24], there is no experimental report on the electronic structure up to date.

Here we systematically investigate the pressure effect on band structures of atomically thin BP (2-13 layer) through Fourier transform infrared (FTIR) spectroscopy. To our surprise, the pressure effect exhibits strong layer-dependence, which strongly deviates from the bulk counterpart. For instance, with the pressure increasing to 2 GPa, the bandgap of 2 L BP is enlarged rather than shrinking, while the bandgap of the bulk is already closed at such pressure. Besides, due to the interlayer interaction, there are multiple optical resonances in few-layer BP, and pressure effect on these optical resonances also varies systematically. Through a tight-binding model in conjunction with a Morse potential, the quantitative evolution of the interlayer overlapping integrals is captured, and the physical mechanism responsible for the layer-dependent pressure effect is unveiled. Our work highlights the critical role played by the tunable interlayer coupling in van der Waals materials.

In our study, few-layer BP was mechanically exfoliated from the bulk crystal onto polydimethylsiloxane (PDMS), then was transferred onto the diamond surface of a diamond anvil cell (DAC), as shown in Fig. 1(a). The layer number and crystal orientation were identified through infrared (IR) extinction spectrum [4,5]. The applied pressure was monitored by the shift of the photoluminescence (PL) peak of a ruby ball placed in the cell. To avoid degradation of few-layer BP, silicone oil was chosen as an inert pressure transmitting medium (PTM) (For details on the experiment procedure, see Methods in Supplemental Materials [25]). It should be noted that PTM usually plays a critical role in pressure experiments on 2D materials. It could compromise the reproducibility of experimental results [29] if the PTM is not optimized. While in our study, the pressure is limited in a moderate range (< 4 GPa), the frequently used PTM is well hydrostatic [30], hence the PTM will not affects our results much and we can treat it as an ideal media.

Figure. 1(b) is the photograph of a thin BP flake in the DAC, containing 3 L and 6 L. Figure. 1(d) shows infrared extinction spectra of this sample under different pressure ranging from 0 GPa (ambient pressure) to 3.22 GPa (the spectra are offset vertically for clarity). Three main peaks can be clearly identified (see Fig. 1(d)), which result from the exciton absorption of 3 L and 6 L BP, labelled as $E_{11}$ (the exciton associated with the first valance band to the first conduction band transition, namely the optical bandgap, see Fig. 1(c)) and $E_{22}$ (the second valance band to the second conduction band). The single particle bandgap has a higher energy than the exciton peak position, with a separation as the exciton binding energy [31,32]. Since the pressure effect on single particle bandgap plays a major role when the pressure is only moderate [33], for simplicity, we neglect the possible change of the exciton binding energy and attribute the band effect to the shift of the exciton peak position. Figure. 1(e) plots the peak positions of $E_{11}$ of 3 L, 6 L, $E_{22}$ of 6 L and the bandgap of bulk BP versus pressure. As we can see, with increasing pressure, the $E_{11}$ of 3 L exhibits a non-monotonic shift, which undergoes redshift from the beginning and then blueshift for pressure above ~

0.6 GPa, in stark contrast to the shift of the bandgap in bulk BP (lower panel of Fig. 1(e)). Meanwhile, $E_{11}$ of 6 L shows a contrasted behavior as well, exhibiting a nonlinear redshift below ~ 1 GPa and then almost no shift from 1 GPa to 2.92 GPa. The different shift of $E_{11}$ in 3 L and 6 L indicates the pressure effect on the bandgap is strongly layer-dependent. Besides, the pressure-induced shift of $E_{22}$ in 6 L is dramatically different from $E_{11}$ as well, exhibiting almost no change below 1 GPa but a blueshift above 1 GPa.

This transition-index dependence of the pressure effect is further verified in an 11 L BP shown in Fig. 2(a), where three optical resonances ($E_{11}$, $E_{22}$, $E_{33}$) can be observed. As shown in Fig. 2(b), $E_{11}$ undergoes redshift nonlinearly with increasing pressure until fully enters the silicone oil absorption region for pressure above ~ 1 GPa. In contrast, $E_{33}$ firstly redshifts below ~ 0.5 GPa and then blueshifts above ~ 0.5 GPa, very similar to $E_{11}$ of 3 L. The pressure-induced shift of $E_{22}$ is in-between $E_{11}$ and $E_{33}$, which redshifts nonlinearly below 1 GPa and then slightly blueshifts, close to the behavior of $E_{11}$ for the 6 L. In fact, this layer- and transition-index-dependent pressure effect is systematic. Figure. S2 shows more data with different layer number ranging from 2 to 13. Besides the pressure-induced peak position shift, $E_{11}$ (also $E_{22}$) of 6 L is barely discernable under 3.22 GPa, as shown in Fig. 1(d), which could result from the direct to indirect bandgap transition [33,34]. However, $E_{11}$ of 3L is still discernible under this pressure, implying pressure-induced direct to indirect bandgap transition also exhibits strong layer-dependence, which deserves further investigation. In this study, we mainly investigate the pressure effect on the electronic structure in the direct bandgap regime. Bearing in mind of larger error bars for extraction of the peak position in higher pressure (especially higher than 2.5 GPa), we only focus on the pressure below 2.5 GPa in the following analysis.

Now we can take a closer look at the mechanism responsible for this layer- and transition-index dependence of the pressure effect. As mentioned earlier, multiple optical resonances in the same few-layer BP are due to the interlayer coupling. The energy separation of them is proportional to the interlayer coupling [35], which can offer us an unambiguous signature to monitor such coupling under pressure. According to previous studies [4,5], the transition energy of optical resonances in *N*-layer BP can be well described through the tight-binding model:

$$E_{nn}^{N}(P) = E_{g0}(P) - \gamma(P)\cos(\frac{n\pi}{N+1}) \qquad (1)$$

where $E_{g0}(P)$ is the monolayer bandgap at pressure *P*, *n* is the subband index, $\gamma(P)$ is the difference of overlapping integrals for conduction band ($\gamma_c$) and valence band ($\gamma_v$), which is proportional to the interlayer hopping parameter ($t^{\perp}$) [35]. From Eq. (1), $\gamma(P)$ can be easily extracted through monitoring the peak separation between optical resonances in the same layer *N* but with different subband indices (*n* and *m*) with the following equation:

$$\gamma(P) = [E_{nn}^{N}(P) - E_{mm}^{N}(P)]/[\cos(\frac{m\pi}{N+1}) - \cos(\frac{n\pi}{N+1})] \qquad (2)$$

Figure. 3(b) plots the relative change of $\gamma$, i.e., $D\gamma = \gamma(P) - \gamma(0)$, versus *P*, extracted from $E_{22}$-$E_{11}$ of 6, 7, 9 L and $E_{33}$-$E_{22}$ of 11 L. The enhancement of $\gamma$ with increasing pressure is unambiguous in spite of the error bar. This is consistent with the expectation, since pressure can decrease the interlayer distance *d* (as shown in Fig. 3(a)). As we can see in Fig. 3(b), the relative change of $\gamma$ can be as large as 0.32 ± 0.08 eV under 1.92 GPa, corresponding to a remarkable change of (18± 4)% ($\gamma(0)$ is 1.76 eV under ambient pressure [5]). According to our previous studies, only ~5% change of interlayer interaction was achieved by biaxial in-plane strain or thermal expansion [36,37]. Apparently, compared to those tuning schemes, pressure is the most efficient way to tune the interlayer interaction in BP. It should be noted that $\gamma(P)$ extracted from samples with different layer number shows little difference with current error bars, suggesting thickness-independent layer distance under pressure. Hence, for simplicity, we assume no layer-dependence for $\gamma(P)$ in the subsequent analysis.

More quantitatively, we can make use of Morse potential which is typically employed to model interatomic forces in molecules and is also applied for the interlayer interaction in TMDCs, as reported in recent studies [38]. Morse potential is written as $U(d) = U_{depth}[1-\exp(-(d-d_{eq})/d_{width})]^2$, where $U_{depth}$ is the depth of the potential, $d_{eq}$ is the equilibrium distance (here can be regarded as the interlayer distance without external pressure), and $d_{width}$ is the width of the potential. Through $P(d) = -\frac{\partial U}{S \partial d}$, where $S$ is the basal area of BP, the pressure as a function of interlayer distance $d$ can be obtained (see Supplemental Information). Now let us revisit $\gamma$, which is mainly determined by the interlayer hoping parameter $t^{\perp}$ (see Fig. 3(a)). Generally, the hopping parameter is scaled as $d^{-2}$ [39]. However, other models are adopted as well. Previous studies reveal that the interlayer hopping parameter can exhibit an exponential decay with $d$ in bilayer graphene [40]. Here we also take the exponential form for $\gamma$, i.e., $\gamma(d) = \gamma_0 e^{-(d-d_{eq})/d_0}$, where $\gamma_0$ is $\gamma$ under 0 GPa and $d_0$ is a characteristic length. Without loss of generality and for simplicity, we assume that $d_0$ is equal to $d_{width}$, as has been done in ref [38]. Then the relationship between $\gamma$ and $P$ reads:

$$\gamma(P) = \frac{\gamma_0}{2} * (1 + \sqrt{1 + \frac{P}{P_{coh}}}) \quad (3)$$

where $P_{coh}$ is termed as cohesive pressure, representing the threshold pressure to be overcome in mechanical exfoliation of BP layers (details in Supplemental Information). Since $\gamma_0$ is known, here only $P_{coh}$ is to be determined. We use Eq. (3) to globally fit $\Delta\gamma$ shown in Fig. 3(b). The overall trend of the model fits well with the experiment data, and $P_{coh}$ is obtained as 2.4±1 GPa, comparable to those obtained in TMDCs [38].

With the obtained relation between the interlayer coupling and the pressure, the pressure effect on intralayer bonds can be accounted as well. According to previous studies [35], due to the absence of the interlayer coupling, monolayer bandgap is mainly determined by two hopping parameters labelled as $t_1^{//}$ and $t_2^{//}$ (see Fig. 3(a)). Besides, our previous studies have shown that the thin BP flake obtained through dry transferring is in good contact with the substrate [36,37]. Moreover, giving the small cross section on the side of a thin BP (noting the force $F=PS$, with $S$ as the side area), it is reasonable to assume that when pressure is not large (as in our case), the thin BP flake always sticks tightly to the diamond surface and the in-plane lattice constants stay the same (the diamond surface has almost no deformation). This is also consistent with the scenario of graphene and quantum wells supported on substrates under pressure [41,42]. Based on this assumption, we know the pressure effect on $E_{g0}$ is mainly due to the change of $t_2^{//}$, since the corresponding bond is out-of-plane, as shown in Fig. 3a. More specifically, with an increasing pressure, the monolayer thickness $D$ (see Fig. 3(a)) decreases. As a result, $t_2^{//}$ is strengthened to induce an increase in $E_{g0}$. This is consistent with theoretical calculations of monolayer BP under normal strain [33,43]. To quantify $E_{g0}$ dependence on $P$, we can make use of Morse potential as well. It should be noted that there is a covalence bond between atoms connected by $t_2^{//}$, and the coupling between them is much stronger than the interlayer coupling, suggesting a larger $P_{\text{coh}}$. When $P \ll P_{\text{coh}}$, as in our case, $E_{g0}$ will change linearly with $P$ (refer to Eq. (3)) as follows:

$$\Delta E_{g0}(P) = aP \tag{4}$$

where $\Delta E_{g0}(P)$ is the relative change of $E_{g0}$, $a$ is the changing rate, which is presumably positive. According to the x-ray diffraction study of few-layer BP [44], the pressure-induced change of the lattice constant along out-of-plane ($D+d$) shows no difference in different thickness BP. Moreover, as mentioned above, the change of the interlayer distance ($d$) also shows no layer-dependence. Hence, it is reasonable to assume that the pressure-induced change of the intralayer height ($D$) has no layer-dependence as well, suggesting $a$ in Eq. (4) is the same for different thickness BP. Based on these assumptions, we can gain some interesting insights. For example, the pressure-induced shift in $E_{11}$ of 3 L, $E_{22}$ of 7 L and $E_{33}$ of 11 L should be the same, since $\cos[n\rho/(N+1)]$ is the same in these optical transitions (see Eq. (1)). Indeed, this is what we have observed in Fig. S3, which further validates our assumptions.

Finally, combining Eq. (1), Eq. (3) and Eq. (4), the pressure-induced shift of the bandgap can be written as:

$$\Delta E_{11}^N(P) = aP - \frac{\gamma_0}{2} * (\sqrt{1 + \frac{p}{p_{coh}}} - 1) * \cos(\frac{1}{N+1}\pi) \qquad (5)$$

which reproduces the experimental results for samples with different thickness up to 50 nm, as shown in Fig. 4. The overall trend of fitting curves agrees well with the experiment data. A global fitting procedure, which excludes the bulk data in Fig. 4 (reasons will be discussed later), gives $a$ of 0.18 ± 0.03 eV/GPa, corresponding to ~ 1% normal strain on monolayer under 1 GPa [43]. Moreover, $P_{coh}$ is 1.4 ± 0.1 GPa, which is consistent with the one extracted from $\Delta\gamma$ in Fig. 3b within error bar. Beside the bandgap, the pressure-induced shift of optical resonances with higher transition-indices ($E_{22}$, $E_{33}$) can be reproduced as well (Shown in Fig. S4). At last, we can clearly see this layer- and transition-index-dependent pressure effect is due to the change of the interlayer interaction $\Delta\gamma$.

To globally fit the data in Fig. 4 with Eq. (5), there is a basic assumption that samples with different thickness have the same lattice constants under a given pressure. Intuitively, this can be satisfied for few-layer and thin films, since they can stick to the diamond surface equally well. However, for even thicker samples, like the bulk one with thickness of 1 μm in Fig. 4, the force induced by the pressure from the side of the flake can be large enough to loose the contact to the diamond surface and the overall pressure effect turns to be truly hydrostatic. In other words, besides the normal strain, there is additional in-plane compressive strain in the bulk BP. As demonstrated previously, such strain also shrinks the bandgap [36]. Therefore, the bandgap of the bulk under hydrostatic pressure decreases faster than the one under normal strain. This explains the dramatic deviation of the bulk sample from others in Fig. 4, even though the bandgap of the bulk is almost the same as the 50 nm thick film if they are under the same condition (e.g. at ambient pressure), as shown in the IR extinction spectrum (See Fig. S5). The bandgap of the bulk redshifts linearly with pressure (shift rate is 0.19 eV /GPa, see Fig. 4), and is ultimately closed at ~ 1.8 GPa, which is fully consistent with previous studies in bulk BP with truly hydrostatic pressure [18]. The different conditions of the bulk and thin film under pressure were further verified by monitoring the phonon frequency through Raman spectroscopy (See Fig. S5). If we regard the bulk sample is under hydrostatic pressure, few-layer and thin films of BP experience non-hydrostatic one, which is more like a uniaxial strain normal to the 2D plane. The substrate plays a very important role and tightly holds the few-layer BP along in-plane directions.

In summary, we have systematically investigated the layer-dependent pressure effect on electronic structures of 2D BP. Through the tight-binding model in conjunction with a Morse potential, we have quantified the evolution of the band gap with pressure, and unveiled that the layer-dependent pressure effect on electronic structures of 2D BP is mainly due to the pressure-induced enhancement of the interlayer coupling. Future studies can be devoted to the possible layer-dependent semiconductor-to-metal transition under higher pressure.

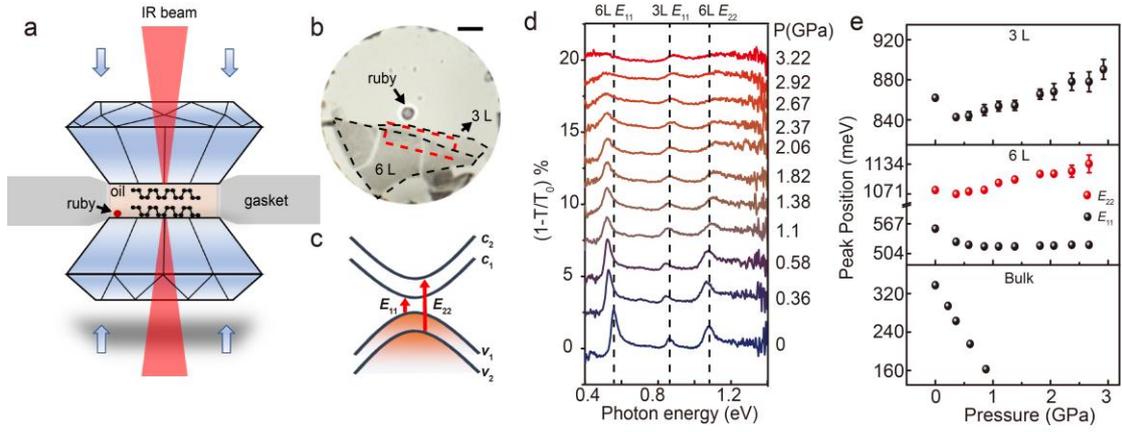

FIG. 1. Pressure effect on electronic structures of few-layer BP. (a) Schematic illustration of the experimental set-up for band structure engineering in few-layer BP through diamond anvil cell (DAC). (b) Optical image of a BP flake which contains 3 L and 6 L in DAC. The region inside the red box is where the IR light shines. Scale bar is 20 μm. (c) Schematic illustration of optical transitions between different subbands in bilayer BP. (d) Infrared extinction spectra of sample shown in Fig. 1(b) under different pressure. The spectra are offset vertically for clarity. (e) Peak position versus pressure for $E_{11}$ of 3 L, $E_{11}$ and $E_{22}$ of 6 L and bandgap of a bulk BP from top to bottom, respectively. Dashed lines guide to the eye.

FIG. 2. Pressure effect on electronic structures of 11 L BP. (a) Infrared extinction spectra of a 11 L BP under different pressure. Silicone oil has strong absorption between 0.35 to 0.37 eV (shaded region). The spectra are offset vertically for clarity. (b) Peak position versus pressure for $E_{11}$, $E_{22}$ and $E_{33}$ of 11 L.

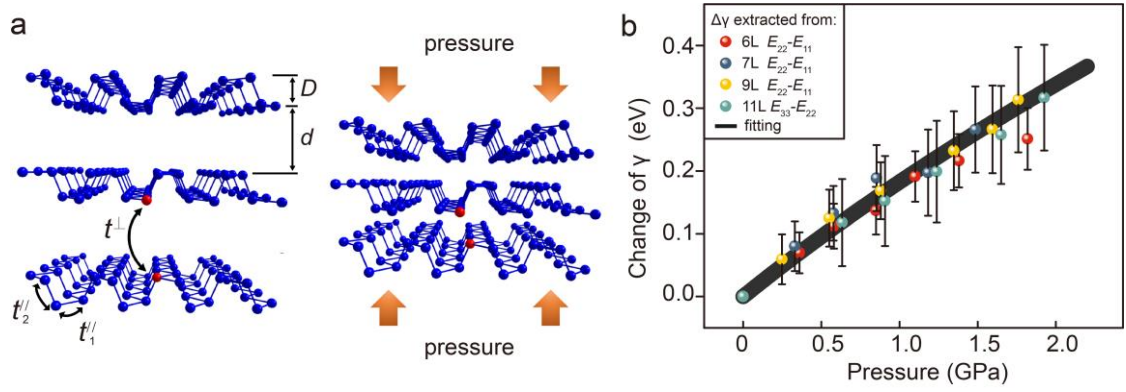

FIG. 3. Pressure effect on the interlayer interaction of few-layer BP. (a) Schematic illustration for the evolution of the atomic structure of tri-layer BP with pressure. (b) The relative changes of γ versus pressure. Dots are data extracted from $E_{22}$-$E_{11}$ of 6 L, 7 L, 9 L and $E_{33}$-$E_{22}$ of 11 L. The solid curve is the fitting based on Eq. (3).

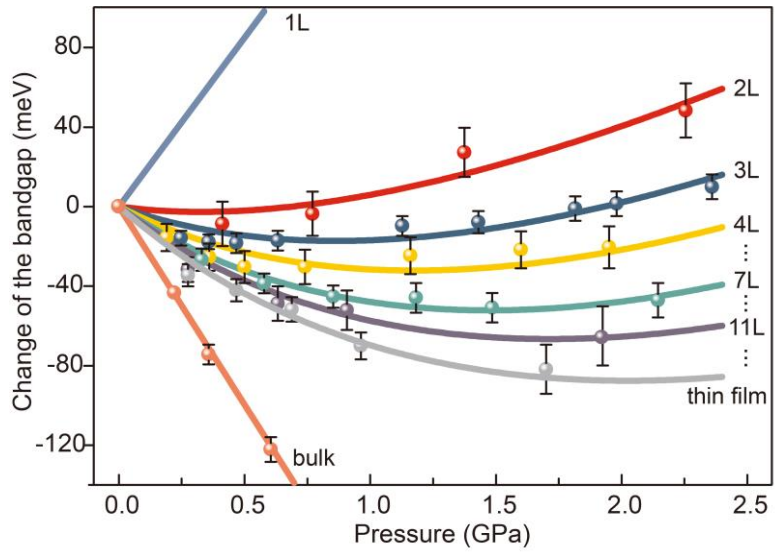

FIG. 4. The relative change of optical bandgaps of BP versus pressure. Dots are experiment data and solid curves are the fitting curves. The thickness of the thin film and bulk shown in the figure are ~ 50 nm and 1 μm, respectively.

**Acknowledgments**

H.Y. is grateful to the financial support from the National Key Research and Development Program of China (Grant Nos. 2017YFA0303504 and 2016YFA0203900), the National Natural Science Foundation of China (Grant Nos. 11874009, 12074085, 11734007), the Natural Science Foundation of Shanghai (Grant No. 20JC1414601) and the Strategic Priority Research Program of Chinese Academy of Sciences (XDB30000000). S.H. acknowledges China Postdoctoral Science Foundation (Grant No. 2020TQ0078). G.Z. acknowledges the financial support from the National Natural Science Foundation of China (Grant No. 11804398), Natural Science Basic Research Program of Shaanxi (Grant No. 2020JQ-105), and the Joint Research Funds of Department of Science & Technology of Shaanxi Province and Northwestern Polytechnical University (Grant No. 2020GXLH-Z-026). This work was partially supported by the National Natural Science Foundation of China (Grants No. 42050203, No. U1530402 and No. U1930401). Y. Lu thank Dr. L.P. Kong of HPSTAR and the staffs from BL01B beamline of National Facility for Protein Science in Shanghai (NFPS) at Shanghai Synchrotron Radiation Facility (SSRF), for assistance with the preliminary HP-IR experiments. Partial experimental research used the resources of SSRF (BL15U1 and BL06B) and the Advanced Light Source (ALS, Beamline 1.4.3), which is a DOE Office of Science User Facility under contract No. DE-AC02-05CH11231.